# Earth's accretion inferred from iron isotopic anomalies of supernova nuclear statistical equilibrium origin

Timo Hopp*,[1], Nicolas Dauphas[1], Fridolin Spitzer[2], Christoph Burkhardt[2], Thorsten Kleine[2]

[1] Origins Laboratory, Department of the Geophysical Sciences and Enrico Fermi Institute, The University of Chicago, 5743 South Ellis Avenue, Chicago, IL 60637, USA

[2] Institut für Planetologie, University of Münster, Wilhelm-Klemm-Str. 10, 48149 Münster, Germany

*Corresponding author: hopp@uchicago.edu

**Abstract**

Nucleosynthetic Fe isotopic anomalies in meteorites may be used to reconstruct the early dynamical evolution of the solar system and to identify the origin and nature of the material that built planets. Using high-precision iron isotopic data of 23 iron meteorites from nine major chemical groups we show that all iron meteorites show the same fundamental dichotomy between *non-carbonaceous* (NC) and a *carbonaceous* (CC) meteorites previously observed for other elements. The Fe isotopic anomalies are predominantly produced by variation in $^{54}$Fe, where all CC iron meteorites are characterized by an excess in $^{54}$Fe relative to NC iron meteorites. This excess in $^{54}$Fe is accompanied by an excess in $^{58}$Ni observed in the same CC meteorite groups. Together, these overabundances of $^{54}$Fe and $^{58}$Ni are produced by *nuclear statistical equilibrium* either in type Ia supernovae or in the Si/S shell of core-collapse supernovae.

The new Fe isotopic data reveal that Earth's mantle plots on or close to correlations defined by Fe, Mo, and Ru isotopic anomalies in iron meteorites, indicating that throughout Earth's accretion, the isotopic composition of its building blocks did not drastically change. While Earth's mantle has a similar Fe isotopic composition to CI chondrites, the latter are clearly distinct from Earth's mantle for other elements (*e.g.,* Cr and Ni) whose delivery to Earth coincided with Fe. The fact that CI chondrites exhibit large Cr and Ni isotopic anomalies relative to Earth's mantle, therefore, demonstrates that CI chondrites are unlikely to have contributed significant Fe to Earth.

## 1. Introduction

Meteorites and planets carry isotopic anomalies of nucleosynthetic origin that were produced in stars that lived before the Sun was born, and escaped homogenization in the interstellar medium and protoplanetary disk (Dauphas and Schauble, 2016). This isotopic heterogeneity may have been inherited from the solar system's parental molecular cloud core (Burkhardt et al., 2019; Dauphas and Schauble, 2016; Jacquet et al., 2019; Nanne et al., 2019) or reflects physical and thermal processing of isotopically anomalous dust in the solar system (*e.g.,* Burkhardt et al., 2011; Dauphas et al., 2002; Trinquier et al., 2009). Regardless of their cause, nucleosynthetic isotopic anomalies provide a powerful tool for reconstructing the evolution of the protoplanetary disk and for establishing genetic ties between terrestrial planets and meteorite parent bodies.

The isotopic compositions of several elements in meteorites display a fundamental dichotomy between *carbonaceous* (CC) and *non-carbonaceous* (NC) materials (for reviews see Bermingham et al., 2020; Kleine et al., 2020; Kruijer et al., 2020). While it was known for some time that carbonaceous chondrites are isotopically distinct from most other meteorites (*e.g.,* Leya et al., 2008; Niederer et al., 1981; Niemeyer, 1988; Rotaru et al., 1992), the first indication of what is now recognized as the NC-CC dichotomy was given by Trinquier et al. (2007), who showed that carbonaceous chondrites have systematically distinct Cr and O isotopic compositions compared to some differentiated meteorites, enstatite chondrites, and ordinary chondrites. Subsequently, Warren (2011) showed that for O, Ti, Cr, and Ni, meteorites always define two isotopically distinct clusters and introduced the term *carbonaceous* and *non-carbonaceous* to distinguish between these two groups. More recent work demonstrated that the NC-CC dichotomy extends to several other elements (*e.g.,* Mo; Budde et al., 2016; Poole et al., 2017; Worsham et al., 2017) and is present in early- and late-formed planetesimals (as represented by iron meteorites and chondrites, respectively) in both reservoirs (Budde et al., 2016; Kruijer et al., 2017). Together, these observations imply that the NC and CC reservoirs were established early in the evolution of the protoplanetary disk, in less than ~1 Ma after the condensation of the first solids in the solar system (refractory Ca,Al-rich inclusions; CAIs), and that they remained spatially isolated at least until accretion of NC chondrite parent bodies at ~2 Ma post CAI formation (Budde et al., 2016; Kruijer et al., 2017). The prolonged spatial separation of the NC and CC reservoirs requires limited mixing and homogenization between the two reservoirs, which may has been accomplished by the early formation of proto-Jupiter (Kruijer et al., 2017) or pressure maxima near the location where Jupiter later formed (*e.g.*, Brasser and Mojzsis, 2020). In addition, the initial separation of the NC and CC reservoirs may also reflect the migration of the snow line that led to two distinct bursts of planetesimal formation at distinct locations in the disk (Lichtenberg et al., 2021).

Despite widespread nucleosynthetic isotopic variations among meteorites and planets, for most elements Earth is isotopically most similar to enstatite chondrites (Dauphas, 2017). This has been interpreted to indicate that Earth's building material was characterized by an enstatite chondrite-like isotopic composition during most of Earth's growth, with only little contribution from CC material (Dauphas, 2017). However, a recent study found that the Fe isotopic composition of CI chondrites are similar to the Earth's mantle, and on this basis argued that the majority of Fe in Earth's mantle derived from CI-like material (Schiller et al., 2020). Moreover, although the same study found two distinct clusters of Fe isotopic anomalies for NC (groups IC and IIIAB) and CC (groups IIC and IVB) iron meteorites, the Fe isotopic composition of CI chondrites appears to be distinct from both, and more similar to NC than CC iron meteorites. Taken at face value, these observations seem to challenge the idea that an isotopic dichotomy for Fe persisted in the protoplanetary disk for millions of years.

In order to better constrain the extent of Fe isotopic variations in the early solar system when the first planetesimals formed, and to evaluate the significance of Fe isotopes for constraining the nature and origin of Earth's building material, we have measured the Fe isotopic compositions of 23 iron meteorites from nine chemical groups belonging to NC (IAB, IC, IIAB, IIIAB, IVA) and CC (IIC, IID, IIIF, IVB) groups. We first evaluate to what extent exposure to galactic-cosmic rays (GCR) during the long residence of iron meteorites in space compromises the quantification of nucleosynthetic Fe isotopic anomalies (Cook et al., 2020). We then use the data to test whether there is an Fe isotopic dichotomy for iron meteorites and to identify the stellar environment where these nucleosynthetic Fe isotopic anomalies initially originated. Finally, we compare the Fe isotopic data with isotopic anomalies of other elements (*i.e.*, Cr, Ni, Mo, Ru) in meteorites and discuss their implications for the evolution of the protoplanetary disk and the identification of Earth's building blocks.

## 2. Samples and analytical methods

### 2.1. Samples and preparation

In this study 23 iron meteorites from nine chemical groups were selected for Fe isotope analyses. Five groups of iron meteorites were previously classified as *non-carbonaceous* (IAB,

IC, IIAB, IIIAB, IVA) and four as *carbonaceous* (IIC, IID, IIIF, IVB) based on their Mo and Ni isotope systematics (Budde et al., 2016; Kruijer et al., 2017; Nanne et al., 2019; Worsham et al., 2019). For 18 iron meteorite samples, the Fe isotopic compositions were analyzed on solution aliquots (~1-2 mg Fe) of digestions that were previously analyzed for their Pt, Mo, Ni, and/or W isotopic compositions (Kruijer et al., 2017; Nanne et al., 2019; Spitzer et al., 2020; Worsham et al., 2019). For five other iron meteorite samples (Toluca, Gibeon, Duchesne, Skookum, Tlacotepec) ~50 mg pieces were cut using a diamond saw, polished with SiC abrasive paper, and cleaned in ethanol. These pieces were then digested in *aqua regia* (3:1 HCl-$HNO_3$) at 120°C for 24 hours on a hot plate. Aliquots of these digestions (~1-2 mg Fe) were processed through iron purification chemistry. Prior to chemical purification, all sample solutions were converted to chloride form and redissolved in 0.25 ml 10 M HCl.

We also analyzed geostandards BHVO-2 and BCR-2 (basalts from Hawaii and Oregon) to assess whether the terrestrial reference material used for reporting Fe isotopic anomalies (IRMM-524a) is an appropriate proxy of the composition of the silicate Earth, as industrial processes used for preparing chemically pure materials can induce spurious isotopic effects (Steele et al., 2011). The basalt samples were digested using HF-$HNO_3$ (2:1) at 150°C on a hot plate for 48 hours followed by several steps of aqua regia, were converted and dissolved in 0.25 ml 10M HCl, and then processed alongside the iron meteorite samples.

### 2.2. Chemical purification

A procedure to purify Fe for measurements of mass-independent isotopic variations that allows efficient separation of Fe from Cu, Ni, Co, and Cr is described by Tang and Dauphas (2012). The sample aliquots are loaded in 0.25 ml 10 M HCl onto 10.5 cm long PFA columns (0.62 cm inner diameter) filled with 3 ml pre-cleaned AG1-X8 (200-400 mesh) anion resin. Iron is fixated on the resin while Ni and other major elements are eluted in 5 ml 10 M HCl. Other possible contaminants (*e.g.*, Cu, Cr) are rinsed off the resin using 30 ml 4 M HCl. Finally, Fe is eluted using 9 ml 0.4 M HCl, dried down, re-dissolved in 10 M HCl, and the chemical purification is repeated using new resin. Finally, the purified Fe is dried down and dissolved in 0.3 M $HNO_3$ for concentration and isotopic measurements. The overall Fe yield is >99% and the total procedural blank is ~70 ng and thus negligible considering that 1-2 mg Fe was purified for each sample. After purification, the interfering elements Cr ($^{54}$Cr on $^{54}$Fe) and Ni ($^{58}$Ni on $^{58}$Fe) were present at low enough levels (Cr/Fe$\leq 1.7\times 10^{-6}$ and Ni/Fe$\leq 2\times 10^{-6}$) to allow accurate Fe isotopic ratio measurements (see Supplementary Material S1; Fig. S1).

*2.3. Isotopic analyses*

Iron isotopic measurements were performed at the University of Chicago using a *Thermo Scientific* Neptune multicollector inductively coupled plasma mass spectrometer (MC-ICP-MS) that was upgraded to Neptune Plus specifications. Ion beams of $^{54}Fe^+$, $^{56}Fe^+$, $^{57}Fe^+$, and $^{58}Fe^+$ were analyzed in static mode on Faraday collectors using a $10^{10}\Omega$ amplifier for $^{56}Fe^+$ and $10^{11}\Omega$ amplifiers for the other Fe isotopes. Possible isobaric interferences from $^{54}Cr^+$ and $^{58}Ni^+$ were measured simultaneously by monitoring $^{53}Cr^+$ and $^{60}Ni^+$ using $10^{12}\Omega$ amplifiers. All Fe ion beams may be affected by molecular interferences from argide ions ($^{40}Ar^{13}C^+$, $^{40}Ar^{14}N^+$, $^{40}Ar^{16}O^+$, $^{40}Ar^{16}O^1H^+$, and $^{40}Ar^{18}O^+$), which need to be resolved for precise isotopic measurements. Therefore, the measurements were made on the flat-topped peak shoulder in either medium-resolution (MR) or high-resolution (HR) mode. We used Ni or Pt sampler and H skimmer cones. Even with Ni cones, little Ni was present and $^{58}Ni^+$ interferences on $^{58}Fe^+$ remained negligible. The main motivation for using Pt cones was an increase in sensitivity and a decrease in the frequency of cone cleaning because the intensity remained more stable.

The purified Fe solutions (10 µg/g in 0.45 M HNO$_3$) were introduced into the MC-ICP-MS at an uptake rate of ~100 µl/min using either a cyclonic glass spray chamber (wet plasma, MR-mode, Pt cones) or an ESI Apex $\Omega$ desolvating nebulizer system (dry plasma, HR-mode, Ni cones) with no auxiliary N$_2$ flow. Typical $^{56}Fe^+$ ion signal intensities were 1.4 nA (wet plasma; MR-mode) to 2 nA (dry plasma; HR-mode), respectively. Each measurement consisted of 25 (HR-mode) or 50 (MR-mode) cycles of 8.369 s each. Sample analyses were bracketed by measurements of the reference material IRMM-524a that has an identical isotopic composition to IRMM-014 (Craddock and Dauphas, 2010). All sample and standard solutions were prepared with the same 0.3 M HNO$_3$ solution. The concentrations of the samples and standards were matched to within ≤2 %, which allows accurate and precise Fe isotopic measurements (see Supplementary Material S1; Fig. S2). On peak zero intensities from a blank solution measured at the beginning of each sequence were subtracted from all individual measurements and a washout time of 210 s was used between all measurements.

Measurements were corrected for instrumental and natural mass fractionation using the exponential law (Dauphas and Schauble, 2016), and internal normalization to either $^{57}Fe/^{56}Fe=0.023095$ or $^{57}Fe/^{54}Fe=0.362549$, the certified ratios of IRMM-014. Mass-independent variations of the isotopic ratios in the samples are expected to be small and we use the µ-notation, which provides the parts-per-million deviation of an isotopic ratio of a sample relative to the mean value of the two bracketing standards,

$$\mu^i\text{Fe}_{(k/j)} = \left[(^i\text{Fe}/^j\text{Fe})_{(k/j),\text{sample}} / (^i\text{Fe}/^j\text{Fe})_{(k/j),\text{IRMM-524a}} - 1\right] \times 10^6, \qquad (1)$$

where the subscript (k/j) indicates the ratio used for internal normalization, in our case either $^{57}\text{Fe}/^{56}\text{Fe}$ or $^{57}\text{Fe}/^{54}\text{Fe}$. Previous studies used the $\varepsilon$-notation (Cook et al., 2020; Cook and Schönbächler, 2017; Dauphas et al., 2008; Völkening and Papanastassiou, 1989) that can easily be converted into $\mu$ by multiplying by 100 (1$\varepsilon$ = 100$\mu$). [see Supplementary Material S2 for a derivation to renormalize published $\varepsilon^{56}\text{Fe}_{(7/4)}$ values to $\mu^{54}\text{Fe}_{(7/6)}$]. The purified solutions are measured $N$ times in a sample-standard bracketing (SSB) scheme and the average $\mu$ values are reported in Table 1. Uncertainties for individual samples are based on repeat measurements ($N$ =10 to 35) of each sample solution using the average value and Student's $t$-value for a two-sided 95% confidence interval (95% c.i.).

We also report mass-dependent isotopic variations, which are given in the $\delta$-notation,

$$\delta^{56}\text{Fe (‰)} = \left[(^{56}\text{Fe}/^{54}\text{Fe})_{\text{sample}} / (^{56}\text{Fe}/^{54}\text{Fe})_{\text{IRMM-524a}} - 1\right] \times 10^3, \qquad (2)$$

allowing us to monitor possible spurious effects on internally normalized isotope ratios introduced by natural mass fractionation (Tang and Dauphas, 2012).

## 3. Results

The Fe isotopic data, together with previously reported Pt isotopic data ($\varepsilon^{196}\text{Pt}_{(8/5)}$; $^{196}\text{Pt}/^{195}\text{Pt}$ ratio internally normalized to a constant $^{198}\text{Pt}/^{195}\text{Pt}$ in parts per 10,000 deviation from a standard) for the same sample digestion or meteorites, are reported in Table 1 and plotted in Figs. 1 and 2. The basaltic geostandards BHVO-2 and BCR-2 give average $\mu^{54}\text{Fe}_{(7/6)}$= 2±2 (95% c.i.) and $\mu^{58}\text{Fe}_{(7/6)}$= 4±6 (95% c.i.), which are normal within uncertainties and agree with previously reported values for terrestrial geostandards (Schiller et al., 2020) (see Supplementary Material S3; Table S1; Fig. S3). We take this value as representative of Earth's mantle because the silicate Earth is thought to be homogeneous from the point of view of isotopic anomalies, and the degree of mass-dependent fractionation in mafic igneous rocks is small (Dauphas et al., 2017 and references therein).

Mass-dependent fractionations in the iron meteorites are small and similar to basalts with $\delta^{56}$Fe values ranging from -0.03 to 0.14 (Table 1). Thus, any spurious isotope effects from non-exponential mass-dependent fractionation should be lower than ~1ppm for both $\mu^{54}$Fe$_{(7/6)}$ and $\mu^{58}$Fe$_{(7/6)}$ (Dauphas and Schauble, 2016; Tang and Dauphas, 2012), which is within the uncertainties of the measurements. The Fe isotopic data of iron meteorites reveal mass-independent variations in $\mu^{54}$Fe$_{(7/6)}$ but no resolvable variations in $\mu^{58}$Fe$_{(7/6)}$. Samples from the four CC iron meteorite groups (IIC, IID, IIIF, IVB) have $\mu^{54}$Fe$_{(7/6)}$ values ranging from +24 to +42. Samples from the NC groups (IAB, IC, IIAB, IIIAB, IVA) have more variable $\mu^{54}$Fe$_{(7/6)}$ values ranging from -5 to +30 (Fig. 1). These results are in good agreement with previously reported data for iron meteorites (see Supplementary Material S3; Table S1; Fig. S3) (Cook et al., 2020; Dauphas et al., 2008; Schiller et al., 2020). For most chemical groups with several measurements (IIC, IID, IIIAB, IVA, IVB), $\chi^2$ tests of homogeneity indicate the individual samples have indistinguishable Fe isotopic compositions (the null hypothesis is that they are indiscernible, and all *p*-values are much higher than 0.05). However, we can resolve small variations within the IC and IIAB iron meteorites when $\mu^{54}$Fe$_{(7/6)}$ is plotted against $\mu^{58}$Fe$_{(7/6)}$.

The isotopic composition of several samples are affected by GCR exposure as monitored by their Pt isotopic compositions (*i.e.*, $\varepsilon^{196}$Pt$_{(8/5)}$) (Table 1). The IC and IIAB iron meteorites samples with the largest within-group $\mu^{54}$Fe$_{(7/6)}$ variations (*i.e.*, Arispe, Bendego, Sikhote Alin, Ainsworth) are also the samples with the longest GCR exposure time and that display the largest Pt isotopic anomalies (transparent symbols in Fig. 1). This suggests that GCR-exposure is responsible for at least some of the observed within-group Fe isotopic variations.

**4. Discussion**

*4.1. Galactic cosmic ray effects and the Fe isotopic dichotomy in iron meteorites*

GCR-induced stable isotope effects in meteorites complicate the use of nucleosynthetic anomalies as genetic tracers. Iron meteorites display longer cosmic ray exposure ages than stony meteorites because they are more likely to survive transit through space due to their material strength (*e.g.,* Herzog and Caffee, 2014). As such, GCR-induced neutron capture effects are more pronounced in iron meteorites and have been reported for various elements, most notably W, Os, and Pt (*e.g.*, Kruijer et al., 2013; Qin et al., 2015). Cook et al. (2020) used model calculations to evaluate the combined GCR exposure effects on Fe, Ni, and Pt isotopes in IAB iron meteorites. In general, Fe isotopes are affected by both neutron capture and spallation. Comparing cosmogenic effects of $\mu^{54}Fe_{(7/6)}$ and $\mu^{58}Fe_{(7/4)}$ with $\epsilon^{196}Pt_{(8/5)}$ is meaningful because the cosmogenic effects for all isotopes involved are thought to be primarily driven by neutron capture effects and they are their own targets, meaning that the chemical composition of the iron meteorite should have little influence on the magnitude and correlation of cosmogenic effects on Fe isotopes. Neutron capture induces net losses of $^{54}Fe$ and $^{56}Fe$, and net gains of $^{57}Fe$ and $^{58}Fe$. Spallation induces a net gain of $^{54}Fe$ and net losses of $^{56}Fe$, $^{57}Fe$, and $^{58}Fe$, but the effects are significantly smaller than those induced by neutron capture (Cook et al., 2020). The calculations of Cook et al. (2020) show that at depth (*e.g.*, 40-100 cm), combined neutron capture and spallation reactions on Fe isotopes in IAB iron meteorites lead to an increase in $\mu^{54}Fe_{(7/6)}$ and a decrease in $\mu^{58}Fe_{(7/4)}$. These authors also argued that (*i*) pre-atmospheric size had little influence on the correlation between GCR-induced shifts in $\mu^{54}Fe_{(7/6)}$ and $\mu^{58}Fe_{(7/4)}$, and (*ii*) because these shifts are mostly induced by neutron capture on Fe itself, the Fe/Ni ratio is not expected to have a significant influence. The predicted overall effects are small and Cook et al. (2020) did not detect any isotopic shift in the IAB iron meteorite dataset outside of the ±12 ppm uncertainty on $\mu^{54}Fe_{(7/6)}$ (see Supplementary Material S3; Table S1).

The data of this study may be used to more completely assess the significance of GCR effects on Fe isotopes in iron meteorites, because several iron meteorites having large GCR effects in the neutron capture dosimeter $\epsilon^{196}Pt_{(8/5)}$ were analyzed. For instance, the samples of this study include Ainsworth (IIAB), which is one of the most strongly irradiated magmatic iron meteorites known. Fig. 2 shows that Fe isotopic anomalies in group IC and IIAB iron meteorites are correlated with their Pt isotopic anomalies. However, the predicted GCR effects are of the same scale as the precision of the isotopic analyses, meaning that these effects can be difficult to resolve. Nevertheless, the correlations between $\mu^{54}Fe_{(7/6)}$ and $\epsilon^{196}Pt_{(8/5)}$ among IC and IIAB iron meteorites are significant, defining slopes of 20±18 and 12±7, respectively, which agree with the GCR model slope for IABs (~13) from Cook et al. (2020) (Fig. 2a,b). By contrast, no significant correlation is observed for IID iron meteorites, but the slope of a linear regression between the IID sample with the lowest $\mu^{54}Fe_{(7/6)}$ (N'Kandhla) and the most irradiated sample (Carbo) of 7±9 is also consistent within error with the expected slope of ~13 and the data plots on the calculated model line (Fig. 2c). As such, these data are not inconsistent with small GCR effects on Fe isotopes in IID iron meteorites.

Assessing the magnitude and extent of nucleosynthetic Fe isotopic variations among iron meteorites requires either correction of GCR effects or exclusion of samples affected by GCR effects from the dataset. Most of the samples of this study display no significant GCR-effects in $\varepsilon^{196}Pt_{(8/5)}$ and based on the GCR-model and the uncertainties of the Fe isotopic data, we, therefore, excluded four samples that have $\varepsilon^{196}Pt_{(8/5)} >0.16$ to calculate low-exposure averages of Fe isotopic anomalies for the different iron meteorite groups (Table 1). The reliability of this criterion was tested by comparing the pre-exposure values of IC and IIAB iron meteorites with the pre-exposure value defined by the intercept of the $\varepsilon^{196}Pt_{(8/5)}$-$\mu^{54}Fe_{(7/6)}$ correlations (Fig. 2; Table 1). Recently, Spitzer et al. (2021) showed that some ungrouped iron meteorites have small nucleosynthetic Pt isotopic anomalies ($\varepsilon^{196}Pt_{(8/5)}$=-0.06±0.01), but this only minimally affects the GCR-correction on $\mu^{54}Fe_{(7/6)}$ values (<0.7ppm). The good agreement of the low-exposure averages and the intercept-derived pre-exposure $\mu^{54}Fe_{(7/6)}$ of the IC and IIAB iron meteorites show that both methods provide reliable and accurate pre-exposure Fe isotopic compositions (Table 1). Additionally, correction of individual samples using the $\varepsilon^{196}Pt_{(8/5)}$-$\mu^{54}Fe_{(7/6)}$ slope defined by IIABs (which is similar to the IAB model) results in similar pre-exposure values but larger uncertainties due to the uncertainty on the slope (see Supplementary Material S4; Table S2). Five samples of this study were not analyzed for their Pt isotopic compositions on the same digestion aliquot, and the absolute GCR effects for these samples are therefore not well constrained. In those cases, we use $\varepsilon^{196}Pt_{(8/5)}$ data for other samples of the same meteorite but because these effects are expected to be influenced by the depth in the pre-atmospheric body, the actual GCR effects in different pieces of these samples may be variable. However, the Fe isotopic composition of these samples (*i*) agree well with other samples from the same group, displaying no significant within group variations (IVAs, IVBs; Table 1), (*ii*) are in very good agreement with literature values (IABs, IVBs; Table S3) (Cook et al., 2020; Schiller et al., 2020), and (*iii*) show similar signatures as other samples from the same group, *i.e.* CC meteorites (IVBs). These samples are, therefore, also included in the discussion.

Fig. 3 displays the average pre-exposure Fe isotopic anomalies of nine iron meteorite groups together with Ni isotopic data for the same groups (compiled in Table S3; all Ni isotopic data relative to NIST SRM986) (Cook et al., 2020; Nanne et al., 2019; Steele et al., 2011; Tang and Dauphas, 2014, 2012). The average pre-exposure $\mu^{54}Fe_{(7/6)}$ values of NC and CC iron meteorites are distinct from each other (Fig. 3a), where a $\chi^2$ test of homogeneity indicates a statistically significant difference between the NC and CC populations (the null hypothesis is that they are indiscernible, and the *p*-value is <<0.05). Furthermore, the iron meteorite groups define two distinct NC-CC clusters in $\mu^{54}Fe_{(7/6)}$-$\mu^{58}Ni_{(2/1)}$ space (Fig. 3b), indicating that the NC-CC dichotomy previously identified for several elements in meteorites extents to the Fe isotopic anomalies of (at least) major iron meteorite groups. The inferred accretion ages for these iron meteorites are <1 Ma after CAI formation (Kruijer et al., 2017; Spitzer et al., 2021), indicating that, like for other elements, the Fe isotopic compositions of the NC and CC reservoirs were established within the first ~1 Ma of solar system formation.

*4.2. Stellar origin of neutron-poor $^{54}$Fe and $^{58}$Ni isotopic anomalies*

Isotopic anomalies in meteorites and their constituents can help constrain the stellar environments that contributed material to the solar system's parental molecular cloud core. However, identifying which isotopes of an element vary is often non-trivial because the small magnitude of isotopic anomalies in bulk meteorites requires internal normalization to a fixed isotope ratio to correct for instrumental and natural mass-dependent fractionation. For example, variation in an internally normalized isotope ratio, such as $\mu^{54}$Fe$_{(7/6)}$, can arise from variation of one or several of the isotopes involved (*e.g.*, $^{54}$Fe, $^{56}$Fe, or $^{57}$Fe).

The data of this study reveal resolvable isotopic anomalies in the pre-exposure $\mu^{54}$Fe$_{(7/6)}$ values, but not in the pre-exposure $\mu^{58}$Fe$_{(7/6)}$ values, which are normal within uncertainties of ~±15 ppm (Fig. 3; Table 1). If the culprit for the variations in $\mu^{54}$Fe$_{(7/6)}$ was solely $^{56}$Fe, then variations in $\mu^{58}$Fe$_{(7/6)}$ of ~-10 would be expected in CC iron meteorites by way of normalization. In contrast, if the variations in $\mu^{54}$Fe$_{(7/6)}$ were solely caused by $^{57}$Fe, then variations in $\mu^{58}$Fe$_{(7/6)}$ of ~-30 would be expected (Fig. 4). The absence of significant variations in $\mu^{58}$Fe$_{(7/6)}$, therefore, suggests that $^{56}$Fe, $^{57}$Fe, and $^{58}$Fe are within uncertainty present in proportions that correspond to the terrestrial standard composition. Although we cannot exclude that nucleosynthetic isotopic variations would mimic mass-dependent fractionation on $^{56}$Fe, $^{57}$Fe, and $^{58}$Fe, the most straightforward explanation is that the observed Fe isotopic variations predominantly reflect anomalies in $^{54}$Fe (Fig. 4). The observation that the neutron-poor nuclide $^{54}$Fe is responsible for the Fe isotopic heterogeneity is consistent with the view that the primary driver for the Ni isotopic variations among bulk meteorite is the neutron-poor nuclide $^{58}$Ni (Steele et al., 2012). Thus, for both, Fe an Ni, CC meteorites reveal excesses in the neutron-poor isotopes (Fig. 3).

Iron and Ni isotopes are primarily produced in massive stars, either during quiescent burning stages or during supernovae explosions of core-collapse (cc-SN; the end point of a massive star when core neutronization is unable the sustain the pressure of the overlying layers) and type Ia (SNIa; explosion of a white dwarf near the Chandrasekar mass by accretion from a companion star). The high Fe/O ratio of the solar system points to 1/3-1/2 of Fe in galactic matter being produced by cc-SN, while the remaining comes from SNIa (Heger et al., 2014). Both $^{54}$Fe and $^{58}$Ni are produced in large amounts in (*i*) SNIa with low density at ignition ($\rho_{ign.}$) (*e.g.,* Iwamoto et al., 1999) and (*ii*) the inner regions of cc-SN. In both cases, explosive Si burning approaches *nuclear statistical equilibrium* (Nomoto et al., 1984), which favors the production of isotopes with near-equal numbers of protons and neutrons (Z=26 and N=28 for $^{54}$Fe; Z= 28 and N=30 for $^{58}$Ni). Radioactive $^{56}$Ni (Z=N=28) and $^{60}$Zn (Z=N=30) are also produced in large amounts during explosive Si burning, and they subsequently decay to $^{56}$Fe and $^{60}$Ni, explaining their high cosmic abundances. Many presolar grain types were produced in Asymptotic Giant Branch (AGB) stars but as discussed by Dauphas et al. (2008), addition of material from AGB stars would create large collateral effects on $^{57}$Fe and $^{58}$Fe, which is not observed. Thus, the most likely origin of the observed $^{54}$Fe variations are SNIa and/or cc-SN.

Further constraints on the origin of the $^{54}$Fe and $^{58}$Ni variations can be gained by testing whether the bulk addition of supernovae (SNe) material or the selective admixture of material from different shells of cc-SNe to material of solar composition can reproduce the $^{58}$Ni and $^{54}$Fe variations among meteorites. The slope of a correlation of isotopic anomalies of two elements *A* and *B* caused by mixing of stellar material to material with solar composition is given by Dauphas et al. (2014),

$$\mu_{k/i}^{j/i} A = \frac{\rho_A^{j/i} - \frac{\omega_{j/i}}{\omega_{k/i}} \times \rho_A^{k/i}}{\rho_B^{q/p} - \frac{\omega_{q/p}}{\omega_{r/p}} \times \rho_B^{r/p}} \times c_{pB}^{iA} \times \mu_{r/p}^{q/p} B, \tag{3}$$

where superscript and subscripts *i, j, k, p, q,* and *r* are isotopes of elements *A* and *B*, respectively, $\rho_E^{j/i} = (^jE/^iE)_{star}/(^jE/^iE)_{sun} - 1$ is the isotopic composition of a stellar source of an element E (*i.e., A* or *B*) normalized to the terrestrial composition, $\omega_{j/i} = \ln(m_{j_E}/m_{i_E})$ where *m* denotes the masses of the isotopes of element E, and $c_{pB}^{iA} = (^iA/^pB)_{star}/(^iA/^pB)_{sun}$ is the curvature coefficient of the mixing relationship between the isotopic anomalies of elements *A* and *B* (Dauphas et al., 2014, 2004). Fig. 5 summarizes the calculated $\mu^{58}$Fe$_{(7/6)}$-$\mu^{54}$Fe$_{(7/6)}$ and $\mu^{64}$Ni$_{(8/1)}$-$\mu^{62}$Ni$_{(8/1)}$ slopes produced by addition of the bulk ejecta from different SNIa and cc-SN nucleosynthesis models to material with solar composition. Steele et al. (2012) argued that only one model output of a SNIa can explain the $\mu^{64}$Ni$_{(8/1)}$-$\mu^{62}$Ni$_{(8/1)}$ slope defined by meteorites. A more recent set of SNIa calculations contains several models that can reproduce the $\mu^{64}$Ni$_{(8/1)}$-$\mu^{62}$Ni$_{(8/1)}$ by addition of bulk SNIa ejecta (Seitenzahl et al., 2013) (Fig. 5). Moreover, bulk addition of several SNIa and cc-SN bulk ejecta can also produce a $\mu^{58}$Fe$_{(7/6)}$-$\mu^{54}$Fe$_{(7/6)}$ slope of 0.29±0.48 (Fig. 5; calculated from iron meteorite data). However, none of the models considered here can simultaneously produce the slopes of $\mu^{58}Fe_{(7/6)}$-$\mu^{54}Fe_{(7/6)}$ and $\mu^{64}Ni_{(8/1)}$-$\mu^{62}Ni_{(8/1)}$ observed among meteorites (Fig. 5).

The calculations highlighted above were performed using bulk SNe ejecta, but many studies have shown that isotopic anomalies are often carried by presolar grains that can sample distinct regions of stars. The inner Si/S zone of cc-SNe can produce $^{58}Ni$ in large amounts and addition of such material can explain the Ni isotopic anomalies observed in meteorites (Steele et al., 2012). Fig. 6 shows profiles of Fe and Ni isotope abundances as a function of interior mass for a 25 M$_\odot$ cc-SN progenitor (Rauscher et al., 2002). As expected, the Si/S shell is characterized by overproduction of the neutron-poor $^{54}Fe$ and $^{58}Ni$ isotopes, but it also produces significant amounts of $^{56}Fe$ and $^{60}Ni$ (Fig. 6a,c). We calculated the slopes of $\mu^{58}Fe_{(7/6)}$-$\mu^{54}Fe_{(7/6)}$ and $\mu^{64}Ni_{(8/1)}$-$\mu^{62}Ni_{(8/1)}$ produced by admixture of material from specific cc-SN zones to material with solar composition and compare them to the slopes expected for $^{54}Fe$ and $^{58}Ni$ variations in meteorites (Fig. 6b,d). The addition of material from the Si/S shells satisfies the predicted $\mu^{58}Fe_{(7/6)}$-$\mu^{54}Fe_{(7/6)}$ and $\mu^{64}Ni_{(8/1)}$-$\mu^{62}Ni_{(8/1)}$ slopes (Fig. 6b,d). While the $\mu^{64}Ni_{(8/1)}$-$\mu^{62}Ni_{(8/1)}$ slope is independent of the stellar masses (15 M$_\odot$ to 40 M$_\odot$) (Rauscher et al., 2002; Steele et al., 2012), the $\mu^{58}Fe_{(7/6)}$-$\mu^{54}Fe_{(7/6)}$ slope is only reproduced in the Si/S zone of cc-SN with M$_\odot$ ≤25 because yields of $^{54}Fe$ decrease and $^{58}Fe$ increase in the inner zones of cc-SNe with higher stellar masses.

In summary, addition of cc-SNe material from the inner Si/S zone can explain the observed $^{54}Fe$ and $^{58}Ni$ variations in meteorites. In contrast, none of the SNIa and cc-SN bulk addition models considered here can simultaneously explain the Fe and Ni isotopic variations among meteorites. However, further work is needed to evaluate the viability of SNIa models, as models of ejecta show considerable heterogeneity that is best captured by performing nucleosynthesis calculations on tracked particles in three-dimensional models (*e.g.*, Seitenzahl et al., 2013). A difficulty with SNIa models is that dust may not condense in ejecta, but that the atoms ejected could still condense on preexisting grains present in the interstellar medium. Further work is therefore needed to assess the contribution of SNIa material to the Fe and Ni isotopic anomalies.

*4.3. Possible origins of the Fe isotopic heterogeneity in the protoplanetary disk*

Possible scenarios for the origin of the isotopic heterogeneity in the protoplanetary disk documented here for Fe are *(i)* selective thermal processing of isotopically distinct phases in an initially homogenized protoplanetary disk (Burkhardt et al., 2012; Dauphas et al., 2002; Trinquier et al., 2009), *(ii)* unmixing of isotopically distinct components in the protoplanetary disk through physical processes such as size sorting (Dauphas et al., 2010; Dauphas et al., 2002; Regelous et al., 2008) or *(iii)* the projection and subsequent processing of the isotopic heterogeneity of the solar system's parental molecular cloud core onto the protoplanetary disk during collapse (Burkhardt et al., 2019; Dauphas and Schauble, 2016; Jacquet et al., 2019; Nanne et al., 2019).

Burkhardt et al. (2019) and Nanne et al. (2019) argued that CAIs represent the isotopic composition of the material responsible for the NC-CC isotopic dichotomy, because the isotopic anomalies of a large number of elements in CC meteorites are offset from NC meteorites towards the isotopic composition of CAIs. If true, CAIs would be expected to have large excesses of $^{54}$Fe (*e.g.*, ~+90 for $\mu^{54}$Fe$_{(7/6)}$ if we assume a simplified linear correlation between $\mu^{54}$Fe$_{(7/6)}$ and $\varepsilon^{54}$Cr among NC, CC, and CAIs). Völkening and Papanastassiou (1989) found large excesses in neutron-rich $^{58}$Fe in FUN-CAIs but no resolvable variations in normal CAIs. Shollenberger et al. (2019) found large negative $\mu^{54}$Fe$_{(7/6)}$ values in mineral separates from a type B CAI but non-FUN bulk CAIs revealed no resolvable variations of individual bulk CAIs relative to CC chondrites. The average $\mu^{54}$Fe$_{(7/6)}$ of 25 bulk CAIs measured by Shollenberger et al. (2019) is 43±17 (95% c.i.), which overlaps with values expected for their host CV and CK chondrites (Schiller et al., 2020). This suggests that the Fe isotopic composition of bulk CAIs is fully or partially overprinted by parent-body aqueous alteration (Shollenberger et al., 2019). Thus, it is currently not possible to evaluate whether material with CAI-like isotopic composition is responsible for the NC-CC Fe isotopic dichotomy observed among iron meteorites.

The Fe isotopic anomalies in iron meteorites do not only display a NC-CC dichotomy but also correlate with isotopic anomalies of Mo and Ru (Fig. 7). The Fe isotopic variations within the NC and/or CC clusters are not well resolved but the clear correlations with the isotopic anomalies of Mo and Ru provide strong evidence that the small $\mu^{54}$Fe$_{(7/6)}$ variations among NC and CC iron meteorites are significant. All NC and some CC chondrites plot on the multi-element isotopic correlations defined by iron meteorites (Fig. 7), and only CI and CO chondrites seem to fall off these correlations (Fig. 7). This and the observation that the majority of chondrite groups plot in the NC-CC clusters defined by iron meteorites (Fig. 8a) suggests that the NC-CC Fe isotopic dichotomy observed for iron meteorites was maintained in the disk at least until the formation time of most chondrite parent bodies.

Earth's mantle plots towards the end of the correlations defined by Fe, Mo, and Ru isotopic anomalies, and furthest away from most CC meteorites (Fig. 7). This contrasts with isotopic anomalies of other iron peak elements (*e.g.*, Cr, Ti) for which Earth's mantle does not have an endmember composition (Warren, 2011). The isotopic variations of Mo and Ru predominantly reflect variations in *s*- and/or *r*-process nuclides (*e.g.,* Burkhardt et al., 2011; Dauphas et al., 2004; Fischer-Gödde et al., 2015; Stephan and Davis, 2021), whose carriers derive from different stellar environments than the carriers of $^{54}$Fe anomalies (Section 4.2). Thus, the observed correlations of Fe, Mo, and Ru isotopic anomalies reveal correlated heterogeneities in material of different stellar origins that are likely hosted in distinct presolar carriers. As such, these correlations are consistent with the view that the large-scale dichotomy and the smaller scale variations within the NC and CC reservoirs represent heterogeneous mixing of two isotopically distinct components with broadly chondritic compositions (Burkhardt et al., 2019; Nanne et al., 2019; Spitzer et al., 2020). These observations are more difficult to account for by thermal processing and/or physical sorting of presolar dust in the nebula, because the isotopic correlations exist for elements across a range of volatilities and of distinct stellar origins.

*4.4. Implications for the accretion of Earth*

The nucleosynthetic isotopic anomalies and the distinct geochemical behavior of elements make it possible to reconstruct the isotopic evolution of the Earth and its building blocks through time (Dauphas, 2017). Based on the observation that CI chondrites have similar Fe isotopic composition to Earth's mantle, Schiller et al. (2020) argued that most of the Fe in Earth's mantle derives from CI-like dust that drifted sunwards through the gaseous disk and was accreted to proto-Earth within the first ~5 Ma of solar system history. Interestingly, CI chondrites also display smaller Mo and Ru *s*-process deficits than other CC groups (Fig. 7) (Burkhardt et al., 2011; Dauphas et al., 2002; Fischer-Gödde and Kleine, 2017). However, while CI chondrites appear to overlap with NC meteorites for Fe and Ru isotopic anomalies, they have *(i)* distinct $\mu^{58}\text{Ni}_{(2/1)}$ values (Fig. 8a) (Steele et al., 2012; Tang and Dauphas, 2014), *(ii)* distinct $\varepsilon^{54}\text{Cr}_{(2/0)}$ (Fig. 8b) (Trinquier et al., 2007), and *(iii)* the same characteristic excess in *r*-process Mo nuclides of CC over NC meteorites (expressed as $\Delta^{95}\text{Mo}$) that is responsible for the Mo isotopic dichotomy (Budde et al., 2019, 2016).

Iron is a moderately siderophile element, and so the Fe isotopic composition of Earth's mantle predominantly represents material accreted during later stages of Earth's accretion. The Fe concentration of Earth's mantle is 6.26 wt.%, whereas the core contains 78.0 to 87.5 wt.% Fe (McDonough and Sun, 1995). This corresponds to an effective metal–silicate partition coefficient $D(\text{Fe})_{\text{M–S}}$ of 12.5 to 14. Iron in Earth is intermediate in its siderophile behavior between Cr ($D_{M-S}$~3) and Ni ($D_{M-S}$~26). Following the approach of Dauphas (2017) we calculated the cumulative density functions (CDFs) for the delivery of Cr, Fe, and Ni atoms during Earth's accretion and their incorporation into Earth's mantle through time (Fig. 9). The CDFs show that ~70% of Cr in Earth's mantle was delivered together with all Fe, and ~70% of Fe was delivered together with all Ni. Thus, the Fe in Earth's mantle inherited a majority of its isotopic signature from the same material that also delivered Cr and Ni. While CI chondrites have similar Fe isotopic compositions to Earth's mantle they display the largest $\varepsilon^{54}Cr_{(2/0)}$ and $\mu^{58}Ni_{(2/1)}$ isotopic anomalies of all carbonaceous chondrites, which are very different from the Earth's mantle value (Fig. 8b) (Steele et al., 2011; Tang and Dauphas, 2014, 2012; Trinquier et al., 2007). Thus, CI chondrites are unlikely to resemble the majority of the material that delivered Fe, Cr, and Ni to Earth's mantle during the accretion of Earth. The similarity of the Fe isotopic composition of CI chondrites with NC meteorites and Earth's mantle, therefore, more likely represents isotopic variability within the CC reservoir, which was subject to admixture of different nucleosynthetic components that coincidently brought the isotopic composition of CI chondrites in the field of NC meteorites for some elements.

Earth's mantle plots on or close to the correlations of Fe, Mo, and Ru isotope anomalies in meteorites (Fig. 7). Ruthenium is a highly siderophile element and its budget in Earth's mantle only represents the last ~0.5% of Earth's accretion whereas Mo records the last ~12% of Earth's growth (Dauphas, 2017). Thus, the fact that the Earth's mantle plots on or close to the correlation lines in the $\mu^{54}Fe_{(7/6)}$-$\varepsilon^{100}Ru_{(9/1)}$ and $\mu^{54}Fe_{(7/6)}$-$\varepsilon^{95}Mo_{(8/6)}$ plots suggests that the material accreted during the last ~60% of accretion of Earth did not significantly change in its isotopic composition (Fig.7). It was previously shown that the Mo isotopic composition of Earth's mantle is intermediate between those of the NC and CC reservoirs, indicating that some addition of CC material during the late stages of Earth accretion is required (Budde et al., 2019; Hopp et al., 2020). However, this minor addition of CC material did not contribute significantly to Earth's mantle Fe budget, which is therefore dominated by NC material. These observations and the predominantly NC-like isotopic composition of Earth are, therefore, consistent with heterogeneous accretion models which proposed that Earth accreted some volatile-rich, and most likely CI-like material towards the end of its growth (Rubie et al., 2011; Schönbächler et al., 2010).

Dauphas (2017) has shown that modeling of the multi-element isotopic evolution of the Earth's mantle during different stages of Earth's accretion suggests that the accreted material always contained large fractions of material with on average enstatite chondrite-like composition, but only minor fractions of carbonaceous chondrite-like material. If the small difference of $\mu^{54}Fe_{(7/6)}$ in enstatite chondrites and the Earth's mantle measured by Schiller et al. (2020) is confirmed, it would call for the presence of an additional component that contributed Fe to Earth that is not yet sampled in meteorites. However, the similarity in Fe isotopic composition between CI and Earth's mantle does not necessarily imply that a significant fraction of Earth's Fe inventory was delivered by CI-like material as in this case collateral isotopic anomalies would have to exist for several other elements.

## 5 Conclusions

High precision Fe isotopic measurements of iron meteorites reveal the presence of isotopic anomalies of both nucleosynthetic and cosmogenic origins. After correction for GCR effects, the nucleosynthetic Fe isotopic anomalies in NC and CC iron meteorites plot in two distinct clusters corresponding to the NC-CC isotopic dichotomy previously observed for other elements. The Fe isotopic anomalies are produced predominantly by variability in the neutron-poor nuclide $^{54}Fe$, consistent with prior observations of variations in the neutron-poor nuclide $^{58}Ni$ among bulk meteorites. For both elements CC iron meteorites are characterized by excesses in neutron-poor isotopes (*i.e.,* $^{54}Fe$ and $^{58}Ni$) relative to NC meteorites and Earth's mantle. These excesses in $^{54}Fe$ and $^{58}Ni$ are best accounted for by the heterogeneous distribution in the early solar system of material produced by *nuclear statistical equilibrium* in the inner regions of cc-SN or possibly SNIa.

The Fe isotopic anomalies of iron meteorites and most chondrite groups correlate with *s*- and/or *r*-process variations in Mo and Ru, providing further evidence that isotopic anomalies in bulk meteorites correlate for elements produced in different stellar environments and having a range of geo- and cosmochemical properties. Earth's mantle plots close to the correlations between Fe, Mo, and Ru isotopic anomalies suggesting that the average isotopic composition of Earth's building blocks did not drastically change throughout the last ~60% of its accretion. A previous study showed that CI chondrites have Fe isotopic compositions similar to Earth's mantle and argued that the majority of Fe in Earth's mantle derives from inward-drifting CI-like dust. However, during accretion and core segregation on Earth, Fe is intermediate in metal affinity between Cr and Ni. Thus, a large fraction of Fe in Earth's mantle was delivered by the same material that delivered Cr and Ni. However, for these two elements CI chondrites are among the most anomalous meteorites relative to Earth, ruling out significant contributions of Cr, Fe, and Ni by accretion of CI-like material.

**Acknowledgments**

Reviews by two anonymous reviewers, and the editorial handling by R. Dasgupta were greatly appreciated. We thank Andy Heard for advice regarding Fe purification and isotopic analysis as well as Emily Worsham for help during the preparation of the sample aliquots. T.H. thanks Quinn Shollenberger for helpful discussions about Fe isotopic analyses and isotopic anomalies in CAIs. We would like to thank the American Museum of Natural History (New York City), the Field Museum of Natural History (Chicago), the National History Museum (London), and the Smithsonian Institution (Washington, DC) for generously providing meteorite samples. **Funding**: This work was supported by NASA grants NNX17AE86G, NNX17AE87G, 80NSSC17K0744, 80NSSC20K0821, and NSF grant EAR-2001098 to N.D. T.K. and C.B. were supported by the Deutsche Forschungsgemeinschaft (DFG, German Research Foundation) – Project-ID 263649064 – TRR 170. This is TRR pub. no. 140.

**CRediT authorship contribution statement**

**Timo Hopp:** Conceptualization, Investigation, Methodology, Validation, Visualization, Writing – original draft. **Nicolas Dauphas**: Conceptualization, Funding acquisition, Resources, Supervision, Writing – review & editing. **Fridolin Spitzer:** Resources, Writing – review & editing. **Christoph Burkhardt:** Funding acquisition, Resources, Writing – review & editing. **Thorsten Kleine:** Funding acquisition, Resources, Writing – review & editing.

**Declaration of competing interest**

The authors declare that they have no known competing financial interests or personal relationships that could have appeared to influence the work reported in this paper.

**Data statement**

All data generated or analyzed during this study are included in this published article (and its supplementary information file).

# Tables

Table 1: Fe and Pt isotopic data of iron meteorites and geostandards.

| Sample | $N^a$ | $\mu^{54}Fe_{(7/6)}{}^b$ | $\mu^{58}Fe_{(7/6)}$ | $\mu^{56}Fe_{(7/4)}$ | $\mu^{58}Fe_{(7/4)}$ | $\delta^{56}Fe^c$ | $\epsilon^{196}Pt_{(8/5)}{}^d$ |
|---|---|---|---|---|---|---|---|
| *Geostandards* | | | | | | | |
| BHVO-2 (HR)$^e$ | 20 | 4±5 | 1±11 | -1±2 | 2±11 | 0.11±0.02 | - |
| Replicate | 15 | 3±5 | 5±7 | -1±2 | 6±6 | 0.14±0.01 | - |
| BCR-2 (HR) | 10 | 3±8 | 1±15 | -1±3 | 2±14 | 0.12±0.01 | - |
| Replicate | 15 | 0±8 | 10±14 | 0±3 | 10±16 | 0.12±0.01 | - |
| *Average* | | 2±2 | 4±6 | -1±1 | 5±5 | | |
| *IAB* | | | | | | | |
| Toluca | 14 | -5±5 | 1±11 | 2±2 | -1±10 | 0.05±0.01 | 0.00±0.09$^f$ |
| *IC* | | | | | | | |
| Arispe | 14 | 16±6 | -8±16 | -5±2 | -2±15 | 0.07±0.03 | 0.36±0.08 |
| Bendego | 14 | 20±8 | -20±10 | -7±3 | -14±8 | 0.14±0.08 | 0.49±0.06 |
| Chihuahua City | 25 | 7±7 | 2±22 | -2±2 | 4±22 | 0.06±0.02 | 0.09±0.07 |
| Mt. Dooling | 15 | 12±7 | -10±12 | -4±2 | -1±13 | 0.05±0.02 | -0.01±0.04 |
| *Low exposure weighted average$^g$* | | 10±5 | -7±11 | -3±1 | 0±11 | | |
| *Intercept of Fe-Pt correlation$^h$* | | 9±5 | -7±11 | -3±2 | 2±11 | | |
| *IIAB* | | | | | | | |
| Ainsworth (HR) | 17 | 30±6 | -20±9 | -10±2 | -11±8 | 0.07±0.02 | 1.09±0.06 |
| Braunau | 15 | 18±6 | -5±14 | -6±2 | 1±14 | 0.08±0.03 | 0.03±0.06 |
| North Chile | 19 | 15±5 | -13±13 | -5±2 | -7±12 | 0.04±0.02 | 0.02±0.03 |
| Sikhote Alin | 15 | 20±6 | -11±13 | -7±2 | -5±12 | 0.11±0.01 | 0.32±0.09 |
| *Low exposure weighted average* | | 16±4 | -9±10 | -6±1 | -4±9 | | |
| *Intercept of Fe-Pt correlation$^h$* | | 16±4 | -9±9 | -6±1 | -3±8 | | |
| *IIC* | | | | | | | - | |
| Perriville | 10 | 42±5 | -3±15 | -14±2 | 10±15 | 0.03±0.05 | 0.05±0.05 |
| Kumerina | 20 | 37±7 | 12±16 | -12±2 | 24±15 | 0.06±0.02 | -0.09±0.03 |
| Wiley (an.) | 15 | 32±5 | 12±11 | -11±2 | 23±11 | 0.04±0.01 | 0.05±0.06 |
| *Weighted average* | | 37±2 | 8±8 | -12±1 | 20±8 | | |
| *IID* | | | | | | | - | |
| Bridgewater | 20 | 28±8 | 17±14 | -9±3 | 26±15 | 0.03±0.03 | -0.01±0.02 |
| Carbo | 35 | 30±5 | -4±9 | -10±2 | 6±9 | 0.10±0.02 | 0.79±0.04 |
| N'Kandhla | 15 | 24±5 | 3±9 | -8±2 | 9±8 | 0.02±0.04 | 0.01±0.02 |
| Rodeo | 15 | 30±6 | 4±18 | -10±2 | 14±18 | 0.07±0.02 | -0.02±0.03 |
| *Low exposure weighted average* | | 27±3 | 7±7 | -9±1 | 13±7 | | |
| *IIIAB* | | | | | | | |
| Cape York (HR) | 14 | 11±8 | 7±17 | -3±3 | 11±17 | 0.03±0.03 | 0.01±0.04 |
| Willamette (HR) | 10 | 9±7 | 8±13 | -3±2 | 16±12 | 0.04±0.02 | -0.05±0.04 |
| *Weighted average* | | 9±5 | 8±10 | -3±2 | 11±10 | | |
| *IIIF* | | | | | | | |
| Clark County (HR) | 10 | 27±7 | 0±14 | -9±2 | 7±14 | 0.03±0.04 | 0.07±0.07 |
| *IVA* | | | | | | | |
| Duchesne$^i$ | 15 | 13±9 | -1±15 | -4±3 | 10±15 | 0.03±0.01 | - |
| Gibeon | 15 | 7±4 | -1±7 | -2±2 | 1±7 | 0.02±0.02 | 0.04±0.08$^f$ |
| *Weighted average* | | 8±4 | -1±6 | -3±2 | 1±6 | | |
| *IVB* | | | | | | | |
| Skookum | 15 | 25±6 | 6±11 | -8±2 | 13±11 | 0.03±0.02 | 0.16±0.05$^f$ |
| Tlacotepec | 15 | 28±6 | 11±12 | -9±2 | 19±11 | 0.02±0.02 | 0.79±0.11$^f$ |
| *Weighted average* | | 26±4 | 8±8 | -9±1 | 16±8 | 0.02±0.02 | |

$^a$ Number of individual analyses.
$^b$ Fe isotopic composition internally normalized to either $^{57}Fe/^{56}Fe=0.023095$ or $^{57}Fe/^{54}Fe=0.362549$ and expressed as µ-notation defined as the parts-per-million deviation of the $^{5x}Fe/^{5x}Fe$ ratio in the sample relative to the two IRMM-524a standard solution bracketing measurements.

[c] Mass-dependent Fe isotopic composition calculated by sample-standard bracketing and given as δ-notation defined parts-per-thousands deviation of the $^{56}$Fe/$^{54}$Fe ratio of the sample relative to the IRMM-524a standard solution.

[d] Pt isotopic composition expressed as the ε-notation (parts-per-10,000 deviation of the $^{196}$Pt/$^{195}$Pt ratio of the sample relative to the standard, internally normalized to $^{198}$Pt/$^{195}$Pt. For all but five samples (Toluca, Duchesne, Gibeon, Skookum, Tlacotepec) the Fe and Pt isotopic composition was analyzed from same the digestion solution. Data from Hunt et al., (2018), Kruijer et al. (2017, 2013), Spitzer et al. (2020), Worsham et al. (2019).

[e] Measurements using the high-resolution mode (HR) setup and all other measurements were made using medium-resolution mode setup as described in section 2.3.

[f] Pt isotopic data for these samples were not measured on the same digestion or know to be from adjacent pieces. The data can be used as a qualitative indicator of GCR exposure. Toluca, Gibeon, and Skookum can be expected to have negligible or low GCR exposure effects whereas Tlacotepec is known to have high GCR exposure effects in its Pt isotopic composition but shows no variation in measured Fe isotopic anomalies of IVB iron meteorites.

[g] Calculated weighted averages (±95% c.i.) using samples with low galactic cosmic ray exposure ($\varepsilon^{196}Pt_{(8/5)}$ ≤0.16) if significant within group variations are observed.

[h] Pre-exposure Fe isotopic composition of IC and IIAB iron meteorites defined by the intercept of the $\mu^{5x}Fe$-$\varepsilon^{196}Pt_{(8/5)}$ correlations. Regression were calculated using the York method.

[i] No Pt isotopic data is available for this sample. The weighted average of the two IVA iron meteorites is similar to the value of the unirradiated sample Gibeon and therefore can be assumed to represents the nucleosynthetic isotopic anomaly of the IVA parent body.

**Figures**

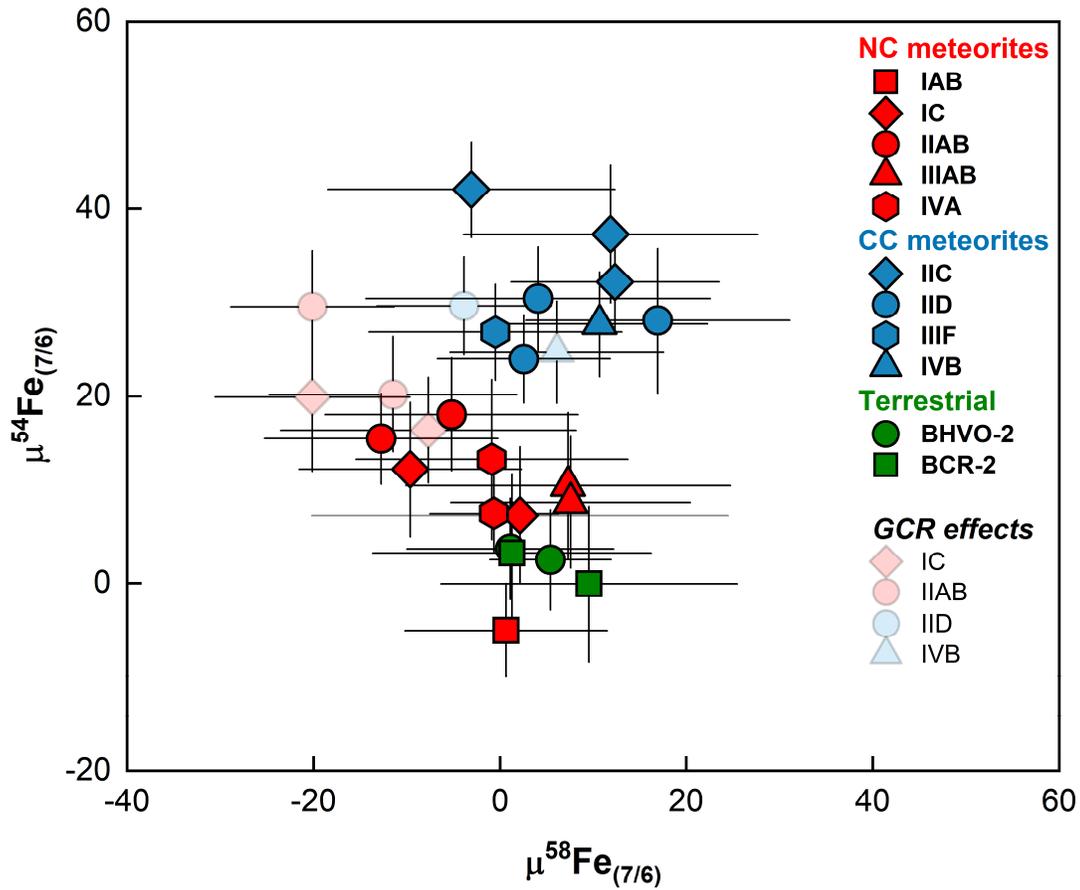

**Fig. 1:** Fe isotopic composition of iron meteorites (red, blue) and geostandards (green). The Fe isotopic data reveal larger $\mu^{54}Fe_{(7/6)}$ values for samples from carbonaceous (CC; blue) relative to samples from non-carbonaceous (NC; red) groups. The $\mu^{58}Fe_{(7/6)}$ values of NC and CC meteorites are indistinguishable. Transparent symbols indicate samples affected by GCR exposure as monitored by their Pt isotopic anomalies (Table 1).

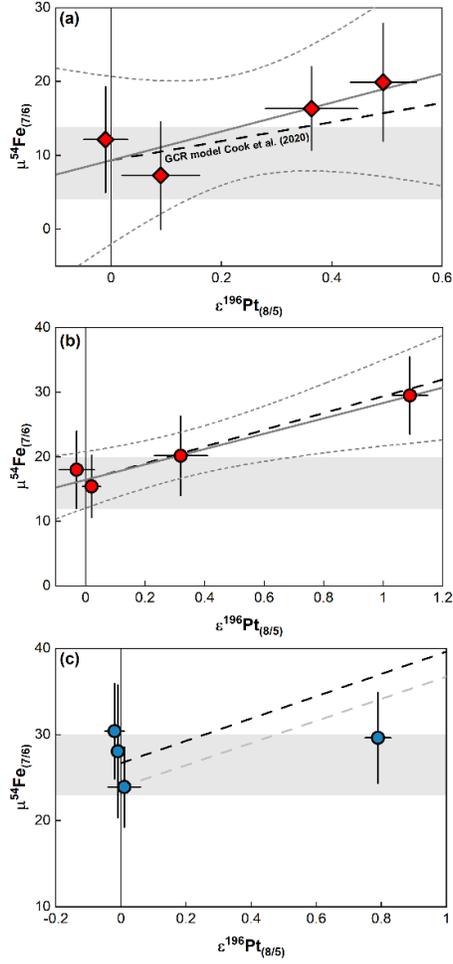

**Fig. 2:** Plots of $\mu^{54}Fe_{(7/6)}$ *versus* $\varepsilon^{196}Pt_{(8/5)}$ for **(a)** IC, **(b)** IIAB, and **(c)** IID iron meteorites. The $\mu^{54}Fe_{(7/6)}$ values of samples within the IC and IIAB iron meteorite groups correlate with $\varepsilon^{196}Pt_{(8/5)}$ **(a,b)** and reveal that exposure to galactic-cosmic rays modified the Fe isotopic composition of some samples. The best fits (solid grey line) and corresponding uncertainties (95% c.i.; dashed grey lines) were calculated using the York method. The correlations agree well with the GCR model calculated for IAB iron meteorites by Cook et al. (2020) (black dashed line). The light grey areas represent the low-exposure average of $\mu^{54}Fe_{(7/6)}$ and corresponding 95% c.i. calculated from the intercepts. **(c)** IID iron meteorites do not show a clear correlation of $\mu^{54}Fe_{(7/6)}$ with $\varepsilon^{196}Pt_{(8/5)}$. The light grey area represent the low-exposure average of $\mu^{54}Fe_{(7/6)}$ and corresponding 95% c.i. calculated from the intercepts. The irradiated sample (Carbo) agrees within uncertainties with the model of Cook et al. (2020) calculated using the unirradiated IID iron meteorite with the lowest $\mu^{54}Fe_{(7/6)}$ (N'Kandhla) as pre-exposure value (grey dashed line). The Pt isotopic compositions are from Spitzer et al. (2020) and compiled in Table 1.

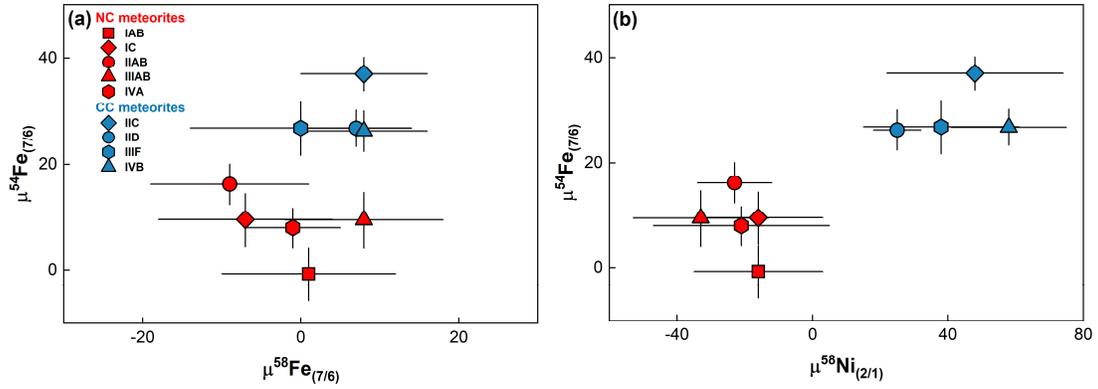

**Fig. 3: (a)** Plot of low-exposure average $\mu^{54}Fe_{(7/6)}$ *versus* $\mu^{58}Fe_{(7/6)}$ of NC (red) and CC (blue) iron meteorite groups (Table 1). The Fe isotopic data reveal two distinct clusters in $\mu^{54}Fe_{(7/6)}$ values of iron meteorites. **(b)** Plot of $\mu^{54}Fe_{(7/6)}$ *versus* $\mu^{58}Ni_{(2/1)}$ shows that CC iron meteorites have larger Fe and Ni isotopic anomalies, and that NC and CC iron meteorites form two distinct clusters. The Ni isotopic compositions of meteorites from the literature are measured relative to NIST SRM 986 and compiled in Table S3 (Cook et al., 2020; Nanne et al., 2019; Steele et al., 2011; Tang and Dauphas, 2014, 2012).

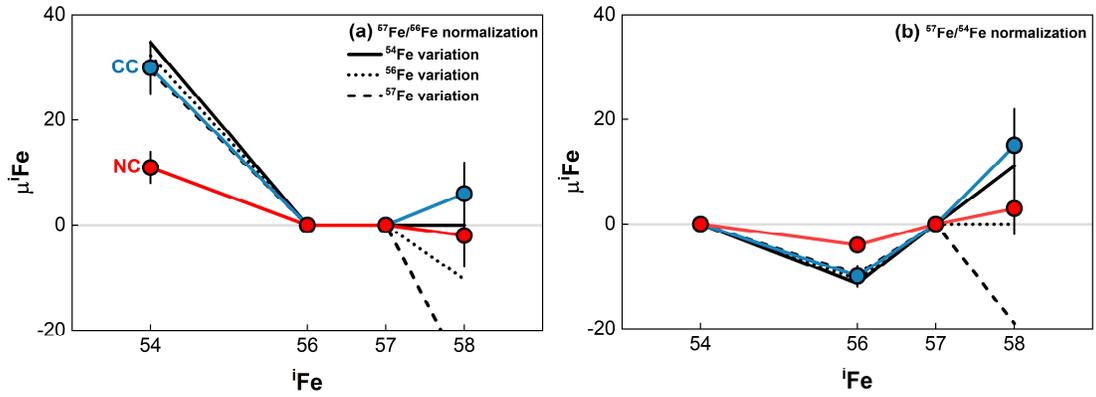

**Fig. 4:** Apparent average nucleosynthetic isotopic anomalies in NC and CC compared to the calculated anomalies that match the $\mu^{54}Fe_{(7/6)}$ and $\mu^{56}Fe_{(7/4)}$ of CC iron meteorites for sole variation in $^{54}Fe$, $^{56}Fe$, and $^{57}Fe$ (black lines) resulting from internal normalization to **(a)** $^{57}Fe/^{56}Fe$ and **(b)** $^{57}Fe/^{54}Fe$. **(a)** When normalized to $^{57}Fe/^{56}Fe$, NC and CC meteorites display resolvable $\mu^{54}Fe_{(7/6)}$ but no resolvable $\mu^{58}Fe_{(7/6)}$ anomalies. Variations in $^{56}Fe$ (dotted line) or $^{57}Fe$ (dashed line) would produce negative $\mu^{58}Fe_{(7/6)}$ anomalies. **(b)** Normalization to $^{57}Fe/^{54}Fe$ results in apparent $\mu^{56}Fe_{(7/4)}$ and $\mu^{58}Fe_{(7/4)}$ anomalies caused by variation in $^{54}Fe$. Variations of $^{56}Fe$ or $^{57}Fe$ would produce no or negative apparent $\mu^{58}Fe_{(7/4)}$ anomalies. The NC and CC averages were calculated from the low-exposure averages of the iron meteorite groups (Table 1).

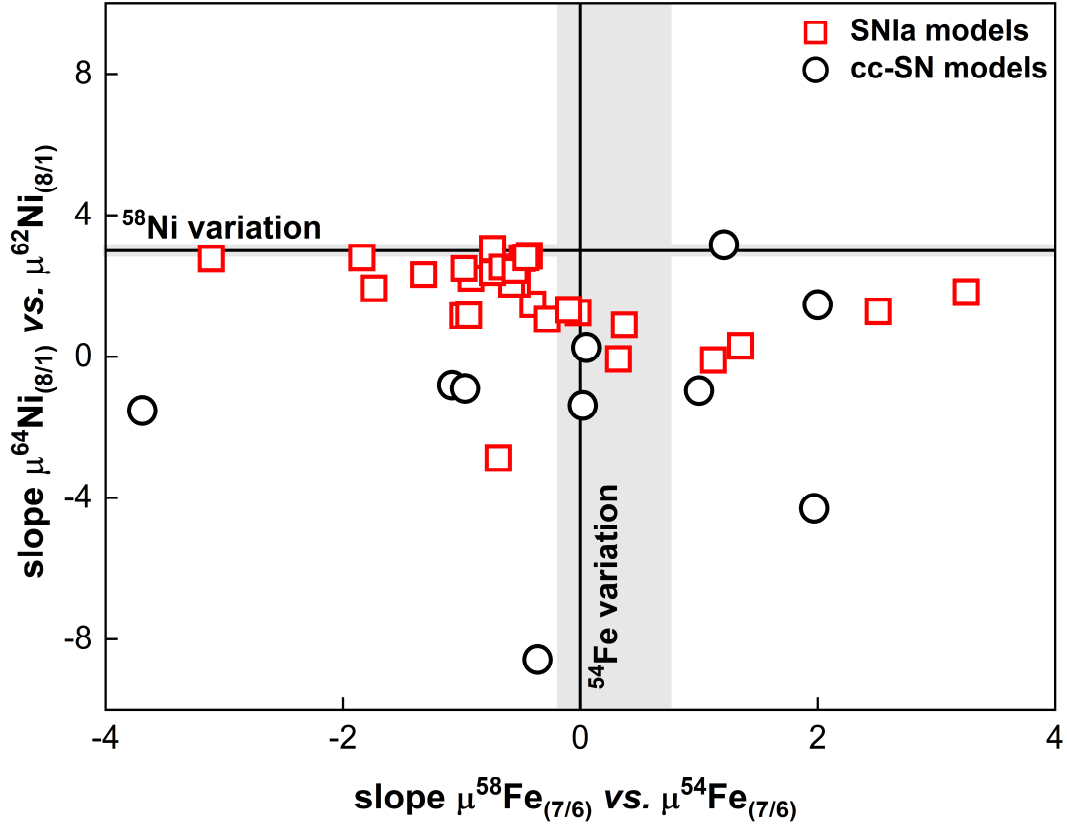

**Fig. 5:** Slopes of $\mu^{64}Ni_{(8/1)}$-$\mu^{62}Ni_{(8/1)}$ *versus* slopes of $\mu^{58}Fe_{(7/6)}$-$\mu^{54}Fe_{(7/6)}$ calculated for admixture of bulk material from various cc-SN and SNIa models to material with solar composition (see section 4.2 for details). The black solid lines define the calculated slopes produced by sole variation of $^{54}$Fe and $^{58}$Ni. The grey areas are defined for Ni by the correlations observed in bulk meteorites (Steele et al., 2012). For $\mu^{58}Fe_{(7/6)}$-$\mu^{54}Fe_{(7/6)}$ the slope (0.29±0.48SE) is defined by the iron meteorite data and calculated using the York method. The slopes of bulk SNe were calculated from data of Iwamoto et al. (1999), Maeda et al. (2010), Rauscher et al. (2002), Seitenzahl et al. (2013), and Woosley (1997).

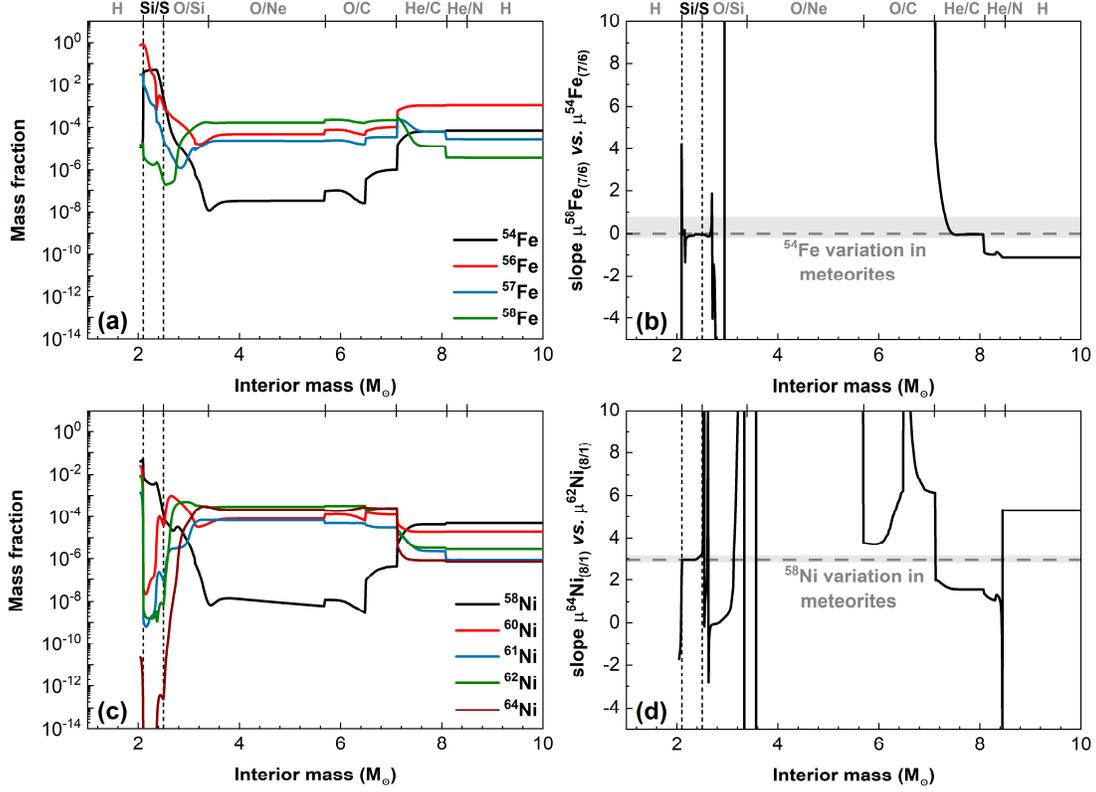

**Fig. 6:** Post-supernova profile of **(a)** Fe isotope abundances and **(c)** Ni isotope abundances as a function of interior mass for a 25 M$_\odot$ cc-SN progenitor computed from zonal yields 25,000 s after core bounce containing the contribution of radioactive progenitors to the stable isotopes (Rauscher et al., 2002). Overabundances of $^{54}$Fe and $^{58}$Ni are produced in the inner Si/S shell (dashed black vertical lines) by nuclear statistical equilibrium. The zones (tick marks on top x-axis) and names are defined by Meyer et al. (1995) and based on the two most abundant elements. **(b,d)** Calculated slopes produced in **(b)** $\mu^{58}$Fe$_{(7/4)}$-$\mu^{54}$Fe$_{(7/6)}$ and **(d)** $\mu^{64}$Ni$_{(8/1)}$-$\mu^{62}$Ni$_{(8/1)}$ space by heterogeneous admixture of small amounts of material from distinct zones to material with solar composition (see section 4.2 for details). The grey dashed lines define the calculated slopes produced by sole variation of $^{54}$Fe and $^{58}$Ni in the solar system. The grey areas are defined for Ni by the correlations observed in bulk meteorites (Steele et al., 2012) and for $\mu^{58}$Fe$_{(7/6)}$-$\mu^{54}$Fe$_{(7/6)}$ calculated from the iron meteorite data of this study (0.29±0.48SE). Only heterogeneous admixing of material from the Si/S zone to material with solar composition can reproduce the observed Fe and Ni slopes of isotopic anomalies in meteorites simultaneously.

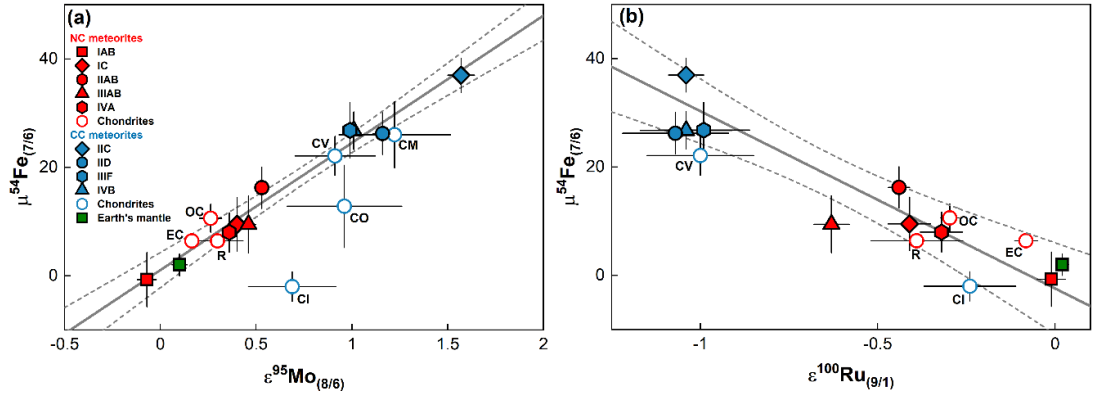

**Fig. 7:** Plots of **(a)** $\mu^{54}Fe_{(7/6)}$ *versus* $\varepsilon^{100}Ru_{(9/1)}$ and **(b)** $\mu^{54}Fe_{(7/6)}$ *versus* $\varepsilon^{95}Mo_{(8/6)}$ of iron meteorites, chondrites, and Earth's mantle. Fe isotopic data of chondrites are from Schiller et al. (2020) and isotopic data of Mo and Ru are compiled in Table S3 (Bermingham et al., 2018; Bermingham and Walker, 2017; Budde et al., 2019, 2018, 2016; Burkhardt et al., 2014, 2011; Fischer-Gödde et al., 2015; Fischer-Gödde and Kleine, 2017; Kruijer et al., 2017; Poole et al., 2017; Render et al., 2017; Spitzer et al., 2020; Worsham et al., 2019, 2017; Yokoyama et al., 2019). The best fits (solid grey line) and corresponding uncertainties (95% c.i.; dashed lines) for the correlation of the isotopic anomalies in iron meteorites were calculated using the York method. Labels correspond to: EC-enstatite chondrites; OC-ordinary chondrites; R-Rumuruti chondrites; CI, CV, CO, and CM are the respective carbonaceous chondrite groups.

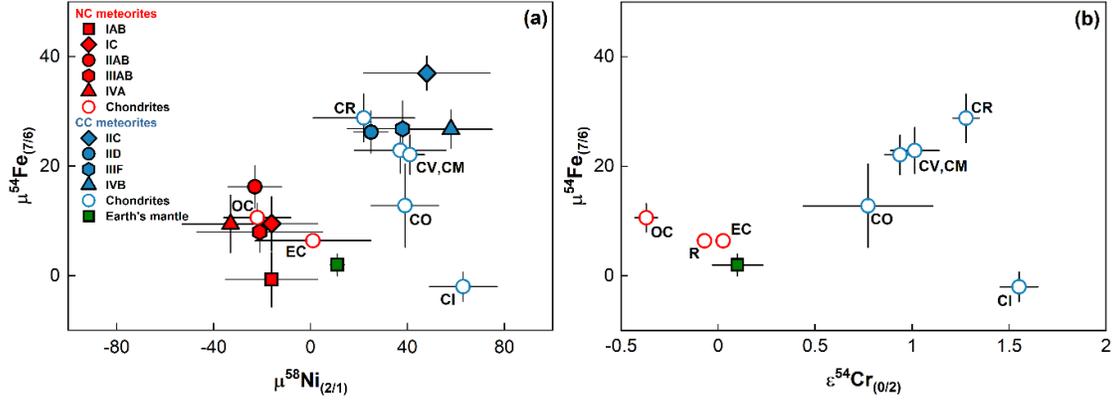

**Fig. 8:** Plots **(a)** $\mu^{54}Fe_{(7/6)}$ *versus* $\mu^{58}Ni_{(2/1)}$ and **(b)** $\mu^{54}Fe_{(7/6)}$ *versus* $\varepsilon^{54}Cr$ of iron meteorites, chondrites, and Earth's mantle. Fe isotopic data of chondrites are from Schiller et al. (2020) and the isotopic composition of Ni and Cr are compiled in Table S3 (Burkhardt et al., 2017; Cook et al., 2020; Göpel et al., 2015; Larsen et al., 2011; Mougel et al., 2018; Nanne et al., 2019; Petitat et al., 2011; Qin et al., 2010; Regelous et al., 2008; Sanborn et al., 2019; Schiller et al., 2014; Schneider et al., 2020; Shukolyukov and Lugmair, 2006; Steele et al., 2012, 2011; Tang and Dauphas, 2014, 2012; Trinquier et al., 2007; Van Kooten et al., 2020, 2016; Zhu et al., 2021). Labels correspond to: EC-enstatite chondrites; OC-ordinary chondrites; R-Rumuruti chondrites; CI, CV, CO, CR and CM are the respective carbonaceous chondrite groups.

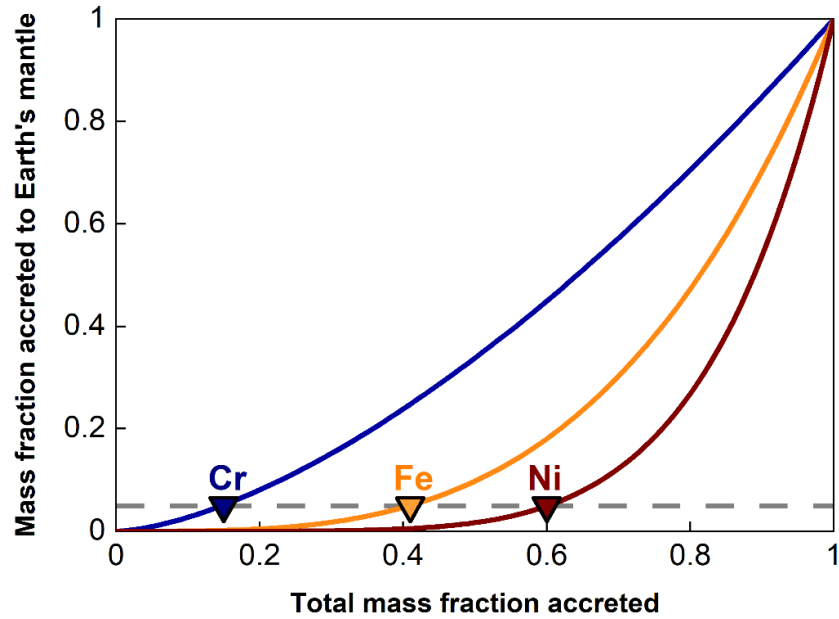

**Fig. 9:** Model of cumulative density functions of Cr, Fe, and Ni for the accretion of elements in Earth's mantle. These curves give the cumulative proportion of atoms that were delivered to Earth's mantle as a function of mass accreted. The curves were calculated using equations given by Dauphas (2017) assuming constant $D(Cr)_{M-S}=3$, $D(Fe)_{M-S}=12.5$, and $D(Ni)_{M-S}=26$. The degree of equilibration of impactor core with target mantle (40%) is assumed to be constant. The triangles mark the position of the fraction that delivered 95% of an element present in the mantle, *e.g.,* 95% of all Fe was delivered by the last ~60% of accretion.